\documentclass[11pt,preprint]{aastex}
\slugcomment{Accepted for the publication in the {\it Astrophysical Journal}}
\def\farcm{\hbox{$.\mkern-4mu^\prime$}}
\def\fsecs{\hbox{$.\mkern-4mu^s$}}
\def\la{\mathrel{\hbox{\rlap{\hbox{\lower4pt\hbox{$\sim$}}}\hbox{$<$}}}}
\def\ga{\mathrel{\hbox{\rlap{\hbox{\lower4pt\hbox{$\sim$}}}\hbox{$>$}}}}

\shortauthors{Park}
\shorttitle{J1601/G330}

\begin{document}

\title{Nonthermal X-Rays from Supernova Remnant G330.2+1.0 and the Characteristics of its Central Compact Object}

\author{Sangwook Park\altaffilmark{1}, Oleg Kargaltsev\altaffilmark{2},
George G. Pavlov\altaffilmark{1}, Koji Mori\altaffilmark{3}, 
Patrick O. Slane\altaffilmark{4}, John P. Hughes\altaffilmark{5}, 
David N. Burrows\altaffilmark{1}, and Gordon P. Garmire\altaffilmark{1}} 

\altaffiltext{1}{Department of Astronomy and Astrophysics, Pennsylvania State
University, 525 Davey Laboratory, University Park, PA 16802; park@astro.psu.edu}
\altaffiltext{2}{Department of Astronomy, University of Florida, Gainesville, FL 32611-2055}
\altaffiltext{3}{Department of Applied Physics, University of Miyazaki, 1-1 Gakuen 
Kibana-dai Nishi, Miyazaki, 889-2192, Japan} 
\altaffiltext{4}{Harvard-Smithsonian Center for Astrophysics, 60 Garden Street,
Cambridge, MA 02138}
\altaffiltext{5}{Department of Physics and Astronomy, Rutgers University,
136 Frelinghuysen Road, Piscataway, NJ 08854-8019}

\begin{abstract}

We present results from our X-ray data analysis of the supernova remnant (SNR) G330.2+1.0 
and its central compact object (CCO), CXOU J160103.1--513353 (J1601 hereafter). Using our 
{\it XMM-Newton} and {\it Chandra} observations, we find that the X-ray spectrum of J1601 
can be described by neutron star atmosphere models ($T^{\infty}$ $\sim$ 2.5--5.5 MK). 
Assuming the distance of $d$ $\sim$ 5 kpc for J1601 as estimated for SNR G330.2+1.0, a small 
emission region of $R$ $\sim$ 0.4--2 km is implied. X-ray pulsations previously suggested by 
{\it Chandra} are not confirmed by the {\it XMM-Newton} data, and are likely not real. 
However, our timing analysis of the {\it XMM-Newton} data is limited by poor photon statistics, 
and thus pulsations with a relatively low amplitude (i.e., an intrinsic pulsed-fraction $<$ 
40\%) cannot be ruled out. Our results indicate that J1601 is a CCO similar to that in the 
Cassiopeia A SNR. X-ray emission from SNR G330.2+1.0 is dominated by power law continuum 
($\Gamma$ $\sim$ 2.1--2.5) which primarily originates from thin filaments along the boundary 
shell. This X-ray spectrum implies synchrotron radiation from shock-accelerated electrons with 
an exponential roll-off frequency $\nu_{\rm rolloff}$ $\sim$ 2--3 $\times$ 10$^{17}$ Hz. 
For the measured widths of the X-ray filaments ($D$ $\sim$ 0.3 pc) and the estimated shock 
velocity ($v_s$ $\sim$ a few $\times$ 10$^3$ km s$^{-1}$), a downstream magnetic field $B$ 
$\sim$ 10--50 $\mu$G is derived. The estimated maximum electron energy $E_{\rm max}$ $\sim$ 
27--38 TeV suggests that G330.2+1.0 is a candidate TeV $\gamma$-ray source. We detect faint 
thermal X-ray emission in G330.2+1.0. We estimate a low preshock density $n_0$ $\sim$ 0.1 
cm$^{-3}$, which suggests a dominant contribution from an inverse Compton mechanism (than
the proton-proton collision) to the prospective $\gamma$-ray emission. Follow-up deep radio, 
X-ray, and $\gamma$-ray observations will be essential to reveal the details of the shock 
parameters and the nature of particle accelerations in this SNR.   

\end{abstract}


\section {\label {sec:intro} INTRODUCTION}

Since nonthermal X-ray synchrotron emission was discovered in portions of SN 1006's 
blast wave shock \citep{koyama95}, several young supernova remnants (SNRs) now show 
such strong particle acceleration sites in the shock front. The detection of nonthermal
X-ray emission in SNRs thus provides an excellent opportunity to study the generation
of high energy cosmic-rays. Based on the archival {\it ASCA} data, Torii et al. (2006) 
discovered that the overall X-ray emission from the Galactic shell-type radio SNR 
G330.2+1.0 shows a featureless spectrum primarily from a power law (PL) continuum 
($\Gamma$ $\sim$ 2.8). G330.2+1.0 is thus one of the rare members of Galactic SNRs 
in which X-ray emission is dominated by nonthermal continuum from synchrotron radiation 
of shock-accelerated relativistic electrons: there are only three other Galactic SNRs 
showing such characteristics, G347.3--0.5 (a. k. a. RX J1713.7--3946, Slane et al. 1999), 
G266.2--1.2 (a. k. a. RX J0852.0--4622 or ``Vela Jr.'', Slane et al. 2001), and G1.9+0.3 
(Reynolds et al. 2008). Torii et al. (2006) noted the general anti-correlation between 
the X-ray and radio intensities in G330.2+1.0, which may suggest multiple populations 
and/or acceleration processes to produce spatially separated X-ray and radio emission. 
The {\it ASCA} study of G330.2+1.0 was, however, limited by the low photon statistics 
($\sim$2000 counts for the entire SNR) and poor angular resolution ($\sim$3$'$ FWHM).  

We observed G330.2+1.0 with {\it Chandra} to perform a detailed imaging and spectral
study of the SNR, which could not be performed with the low angular resolution detectors 
on board {\it ASCA} \citep{park06}. In our initial work on the {\it Chandra} data, we
discovered a candidate neutron star (CXOU J160103.1--513353, J1601 hereafter) at 
the center of the SNR \citep{park06}. The featureless X-ray spectrum of J1601 is
well-described by a black-body (BB) model with $T$ = 5.7 MK. The high foreground 
column ($N_{\rm H}$ $\sim$ 2.5 $\times$ 10$^{22}$ cm$^{-2}$) is consistent with that
for SNR G330.2+1.0, supporting their spatial association. Assuming the distance
$d$ $\sim$ 5 kpc to G330.2+1.0 as estimated by the H{\small I} absorption \citep{mg01}, 
a small area ($R$ $\sim$ 0.4 km) for the BB region is implied. The observed X-ray 
luminosity is $L_{\rm X}$ $\sim$ 1 $\times$ 10$^{33}$ ergs s$^{-1}$ in the 1--10 keV band. 
No counterpart is found in the optical, IR, and radio bands, and a large X-ray-to-optical 
flux ratio (${f_{{\rm 1-10~keV}}}/{f_{\rm V}}$ $>$ 9) is estimated. There is no evidence 
for long-term time variability (in the 1-7 keV band) up to $\sim$10 hr time scales in the 
light curve of J1601. All these characteristics are typical for the peculiar manifestation 
of neutron stars found at the center of several SNRs, dubbed ``Centeral Compact Objects (CCOs)'' 
\citep{pav02}. A particularly intriguing aspect of J1601 is its possible pulsations \citep{park06}. 
Because of the low photon statistics ($\sim$600 counts) and the long frame-time (3.24 s) 
of the {\it Chandra} data, the detection of the periodicity ($P$ = 7.48 s) was not conclusive 
($\sim$2$\sigma$ significance). 

We have recently performed a follow-up observation of G330.2+1.0 and J1601 with {\it XMM-Newton}
to compensate the low photon statistics and poor time resolution of the {\it Chandra} data.
When combined with the high angular resolution {\it Chandra} data, the good time resolution 
and large collecting area of {\it XMM-Newton} can help reveal the detailed nature of the SNR 
and the CCO. We report here the results from our analysis of G330.2+1.0 and J1601 using our 
{\it XMM-Newton} and {\it Chandra} data. Unfortunately, the {\it XMM-Newton} data are 
significantly contaminated by flaring background. Nonetheless, using the available data 
({\it Chandra} + {\it XMM-Newton}), we derive some fundamental properties of J1601 and 
G330.2+1.0. In \S~\ref{sec:obs}, we describe the observations and the data reduction. X-ray 
spectral and timing analyses of J1601 are presented in \S~\ref{sec:cco_spec} and 
\S~\ref{sec:cco_time}, respectively. We present the spectral analysis of G330.2+1.0 in 
\S~\ref{sec:snr}. In \S~\ref{sec:disc}, we discuss implications on the results from data 
analysis of the CCO and SNR. Finally, a summary and conclusions are presented in \S~\ref{sec:sum}. 

\section{\label{sec:obs} OBSERVATIONS \& DATA REDUCTION}

We observed G330.2+1.0/J1601 with the European Photon Imaging Camera (EPIC) on board
{\it XMM-Newton} Observatory on 2008-03-20 (ObsID 0500300101). The pointing (RA[J2000] 
= 16$^h$ 01$^m$ 3$\fsecs$14, Dec[J2000] = --51$^{\circ}$ 33$^{\prime}$ 53$\farcs$6) 
is to J1601 which is positioned at the center of the nearly circular X-ray shell of SNR
G330.2+1.0. We chose the small-window mode (4$\farcm$4 $\times$ 4$\farcm$3 field of view 
[FOV] and 6 ms time resolution) for the EPIC pn to search for pulsations of J1601. 
We chose the full-window mode ($\sim$30$'$ diameter FOV) for the EPIC MOS detectors to
study the entire SNR. The medium filter was used for all detectors. We reduced the 
data using the Science Analysis System (SAS) software package v7.1.0. 

Our {\it XMM-Newton} observations of G330.2+1.0/J1601 were significantly contaminated by 
flaring particle background. We removed time bins in which the overall count rate is 2$\sigma$ 
(the pn) or 3$\sigma$ (the MOS1 and MOS2) above the mean value for time intervals unaffected 
by flaring background. Time intervals including a considerable contamination by the flaring 
background ($\ga$ 50\% above the average quiescent rate) were eliminated by these time-filters. 
After the time filtering, 26, 31, and 33 ks exposures for the pn\footnote{The $\sim$30\% 
deadtime-corrected exposure for the small window mode of the pn is 18.3 ks.}, MOS1, and MOS2, 
respectively, are available for further data analysis, which is $\sim$40--45\% of the total 
exposure. We then reduced the data following the standard screening of event pattern (PATTERN 
$\leq$ 12 for the MOS1/2 and PATTERN $\leq$ 4 for the pn) and hot pixels (FLAG = 0). (For the 
timing analysis of J1601, we used a longer exposure while choosing a smaller aperture and 
more strict event pattern criteria as described in \S~\ref{sec:cco_time}.) There are stable 
components of instrumental background in the EPIC detectors. The primary components that 
could affect this work are Al-K ($E$ $\sim$ 1.5 keV) and Si-K ($E$ $\sim$ 1.7 keV) 
fluorescence lines due to the interactions of high energy particles with the structure 
surrounding the detectors and the detectors themselves\footnote{{\it XMM-Newton} Users' 
Handbook, \S~3.3.7.2.}. We removed these events from our image analysis by excluding
narrow energy bands centered on these lines. Our background-subtracted source spectra show 
little evidence of these lines. Thus, we believe that the impact of contamination from this 
instrumental background on our EPIC data analysis is negligible.

Because of the severe contamination by the flaring background, photon statistics of the
filtered {\it XMM-Newton} data are significantly lower than originally intended. Thus, 
in addition to the {\it XMM-Newton} data, we use the {\it Chandra} data (ObsID 6687) for 
the spectral analysis to improve overall photon statistics. The high angular resolution of 
{\it Chandra} data is also essential to measure the widths of the thin X-ray filaments of 
G330.2+1.0.  We performed the {\it Chandra} observation of G330.2+1.0 with the I-array of 
the Advanced CCD Imaging Spectrometer (ACIS) on 2006-05-22 as part of the Guaranteed Time 
Observations program. The effective exposure after the data screening is $\sim$50 ks, and 
thus photon statistics for the SNR and CCO in the ACIS data are similar to those obtained 
by the EPIC MOS1+MOS2 data. The details of the {\it Chandra} observation and data reduction 
are described by Park et al. (2006).  

\section{\label{sec:cco_spec} X-Ray Spectrum of the Central Compact Object}

We extracted the spectrum of J1601 ($\sim$530, 290, and 310 counts in the 0.5--10 keV band
for the pn, MOS1, and MOS2, respectively) from a circular region with a radius of 
15$^{\prime\prime}$.  The background spectrum was estimated from two nearby source-free 
regions with a radius of 30$^{\prime\prime}$. The background counts contribute $\sim$15\% 
to the pn spectrum and $\sim$9\% to the MOS spectra. The background-subtracted, 
deadtime-corrected count rates (in the 15$^{\prime\prime}$ radius aperture) are $\sim$0.025 
(pn) and $\sim$0.009 counts s$^{-1}$ (MOS1/2). The {\it Chandra} spectrum of J1601 was 
extracted from an $\sim$2$^{\prime\prime}$ circular region. The background spectrum was 
extracted from the surrounding annular region with the inner and outer radii of 
4$^{\prime\prime}$ and 15$^{\prime\prime}$, respectively \citep{park06}. The 
background-subtracted ACIS count rate is $\sim$0.012 counts s$^{-1}$. The total source 
counts combining all the {\it XMM-Newton} EPIC and {\it Chandra} ACIS data are $\sim$1700 
counts, which is about three times higher than those used in the previous work. Each source 
spectrum was binned to contain a minimum of 20 counts per energy bin. 

We simultaneously fit four spectra of J1601 obtained by the {\it XMM-Newton} pn, MOS1, MOS2, 
and the {\it Chandra} ACIS. Initially we fit the spectrum with a BB model. The best-fit BB 
temperature and the absorbing column ($T_{\rm BB}$ = 5.6$^{+0.3}_{-0.4}$ MK, $N_{\rm H}$ = 
2.46$^{+0.38}_{-0.35}$ $\times$ 10$^{22}$ cm$^{-2}$, $\chi^{2}$/$\nu$ = 73.4/74, errors are
at 90\% confidence level [C.L.], hereafter) are consistent with those by Park et al. (2006). 
The implied emitting area is small ($R$ $\sim$ 0.44 $d_5$ km, where $d_5$ is the distance to 
the CCO in units of 5 kpc), which is also in agreement with the previous work. The observed 
flux ($f_{\rm X}$ $\sim$ 1.22 $\times$ 10$^{-13}$ ergs cm$^{-2}$ s$^{-1}$ in the 1--10 keV 
band) is consistent with the previous {\it Chandra} results as well. Although a PL model 
may also fit the data, a very steep photon index ($\Gamma$ = 5.6$^{+0.5}_{-0.4}$) and a high 
$N_{\rm H}$ = 5.4$\pm$0.6 $\times$ 10$^{22}$ cm$^{-2}$ ($\chi^{2}$/${\nu}$ = 93.6/74) 
are implied (The PL fit is not acceptable with $\chi^2_{\nu}$ $>$ 2, when $N_{\rm H}$ is 
fixed at 2.5--3.0 $\times$ 10$^{22}$ cm$^{-2}$). This PL shape is too soft for typical 
synchrotron emission from the neutron star's magnetosphere, and the fit is statistically 
worse than that by the BB model. Thus, we conclude that X-ray emission of J1601 is consistent 
with a BB spectrum. Using only {\it XMM-Newton} data, we estimate the {\it same} flux
($f_{\rm 1-10 keV}$ $\sim$ 1.25 $\times$ 10$^{-13}$ ergs cm$^{-2}$ s$^{-1}$), which 
indicates that flux variations in the two years between the {\it Chandra} (2006-05-22) and 
{\it XMM-Newton} (2008-03-20) observations are negligible ($\la$5\%). 

While the BB model can fit the overall X-ray spectrum of J1601, it may not be physically
adequate to describe thermal emission from a neutron star. A neutron star is not a
perfect BB, and likely has an atmosphere whose emission from the outermost H-layer
may dominate the observed spectrum (e.g., Pavlov \& Zavlin 2000). The observed spectrum of 
the thermal radiation from a neutron star's surface is substantially affected by the properties
of its atmosphere such as chemical composition, magnetic field, gravity, and the energy-dependent 
opacities (Pavlov et al. 1995 and references therein). A typical observational effect may be a 
higher temperature and a smaller emitting area than the ``true'' values when the spectrum 
is fitted by a simple BB model (e.g., Zavlin et al. 1998; Pavlov et al. 2000). Therefore, 
taking advantage of the improved photon statistics in the {\it XMM-Newton} + {\it Chandra} 
data, we fit the observed X-ray spectrum of J1601 with a hydrogen neutron star atmosphere 
model (NSA model in XSPEC, Pavlov et al. 1995; Zavlin et al. 1996). 

First, we fit the spectrum of J1601 with a single NSA model. The magnetic fields of CCOs may 
be significantly lower or higher than the ``canonical'' magnetic field of a pulsar, $B=10^{12}$ 
G (e.g., Pavlov \& Zavlin 2000; Bignami et al. 2003). Based on the lack of a strong absorption
feature in the observed spectrum of J1601, which would be interpreted as an electron cyclotron 
line in an NSA spectrum, we can only exclude fields in the range of $\sim (2-8)\times 10^{11}$ 
G. We thus fit our spectrum using all three magnetic field values for which the NSA models 
are available in XSPEC: $B=0$ (applicable for low fields, $B\lesssim 10^{10}$ G), $10^{12}$, and
$10^{13}$ G. We fix the neutron star mass and radius at the canonical values $M_{\rm ns} = 1.4 
M_\odot$ and $R_{\rm ns}=10$ km, which correspond to the gravitational redshift parameter 
$g_{\rm r}=(1-2GM_{\rm ns}/R_{\rm ns}c^2)^{1/2}= 0.766$, and vary the effective temperature, 
distance, and $N_{\rm H}$. The fits are statistically acceptable for all three magnetic 
field values. For $B=0$ and $10^{13}$ G, the redshifted best-fit effective temperatures,
$T_{\rm eff}^\infty = 2.6$ and 3.6 MK, respectively, are lower than the $T_{\rm BB}$, while 
$T_{\rm eff}^\infty=5.7$ MK for $B = 10^{12}$ G is about the same as the BB temperature\footnote{The 
higher $T_{\rm eff}$ and the correspondingly larger distance in the $B=10^{12}$ G NSA fit are caused 
by the fact that the high-energy part of the observed spectrum coincides with the low-energy wing 
of the the gravitationally redshifted electron cyclotron feature, centered at $\approx 9$ keV. 
Therefore, the high-energy tail of the X-ray spectrum is much softer than those in the models with 
very low or very high magnetic fields. Although we cannot exclude this field value based on the 
observational data available, we note that the models with slightly lower fields ($B$ $\sim$ 2--8
$\times$ 10$^{11}$ G) would not fit the data, while the models with higher fields (e.g., 
$B\gtrsim 2\times 10^{12}$ G) would yield essentially the same results as the $B=10^{13}$ G model.}.
The best-fit $N_{\rm H}$ values, 3.1, 2.4, and 2.7 $\times 10^{22}$ cm$^{-2}$, for $B=0$, $10^{12}$, 
and $10^{13}$ G,  respectively, are consistent with that for SNR G330.2+1.0 ($N_{\rm H}\sim 2.5$--$3
\times 10^{22}$ cm$^{-2}$). The best-fit distances, 24, 169, and 55 kpc, for $B=0$, $10^{12}$, and 
$10^{13}$ G, respectively, are unreasonably large for a Galactic object. To reconcile them with 
the distance to the SNR ($d\sim 5$ kpc; McClure-Griffiths et al.\ 2001), we have to assume that 
the sizes of the X-ray emitting areas are much smaller than $R_{\rm ns}=10$ km ($R\sim 2$, 0.3, and 
0.9 km for $B$ = 0, 10$^{12}$, and 10$^{13}$ G, respectively). Thus, although the fits with the 
$B=0$ and $10^{13}$ G NSA models yield lower temperature and larger sizes than the BB fit, the 
observed emission cannot be interpreted as emitted from the entire neutron star surface.

Based on these results, it is natural to assume that the X-ray emission in J1601 originates 
from small hot spots on the neutron star's surface, such as suggested for other CCOs (e.g., 
the CCO in Cassiopeia A, Pavlov et al. 2000). In this scenario, the observed thermal X-ray 
emission consists of two characteristic components: the hot component from a small region(s) 
and the cool emission from the rest of the stellar surface (e.g., Pavlov et al. 2000). Therefore, 
we fit the observed spectrum of J1601 with two-component NSA models, assuming $B$ = 0, 10$^{12}$, 
or 10$^{13}$ G, the same for both components. For the soft component, we fix the distance to the 
CCO and the size of the emitting region at $d$ = 5 kpc and $R$ = 10 km, respectively, while 
varying the surface temperature. For the hard component, both the distance and surface temperature 
are varied. The foreground column $N_{\rm H}$ is tied common for both components, and then 
is fitted. The results are summarized in Table~\ref{tbl:tab1}. The X-ray spectrum of J1601 
and the best-fit two-component NSA model with a high magnetic field ($B$ = 10$^{13}$ G) 
are presented in Fig.~\ref{fig:fig1}. We note that, although the additional soft component 
is statistically not required, implying only an upper limit on the observed flux (e.g., 
$f_{\rm 1-10 keV}$ $<$ 5 $\times$ 10$^{-14}$ ergs cm$^{-2}$ s$^{-1}$ at 90\% C.L.), 
the two-component model likely represents a physically more realistic picture than the 
one-component model to account for the small hot region implied by the single NSA model. 
Therefore, we hereafter discuss the spectral nature of J1601 based on the two-component 
NSA model fits.

\section{\label{sec:cco_time} Search for Pulsations from the Central Compact Object}

Park et al. (2006) searched for X-ray pulsations from J1601 in the 50 ks {\it  Chandra} 
ACIS observation (3.24 s time resolution)  and reported a marginally significant (at a 
$\approx2\sigma$ level) periodic signal  ($P$ = 7.48 s, a pulsed fraction $f_{rm p}$ 
$\sim$ 30\%). One of the goals of our follow-up {\it XMM-Newton} observation was to test 
the significance of the previously reported period candidate, and to search for periodicity 
outside the frequency range accessible with {\it Chandra} data (The EPIC pn in the 
small-window mode provides much better 6 ms resolution). However, as discussed in 
\S~\ref{sec:obs}, the flaring background hampered our timing analysis of the EPIC pn data. 
For the timing analysis, we performed the data reduction following the methods described 
in \S~\ref{sec:obs}, except that (1) we used 44 ks of uninterrupted pn data after removing 
major flares, (2) we extracted photons from a smaller circular aperture (8$\farcs$4 in radius) 
than the standard pointlike source extraction area for the EPIC (15$^{\prime\prime}$ in radius), 
and (3) we applied a stricter screening by selecting  events with PATTERN = 0. After these
data reduction, we obtained a total of 761 photons (including $\approx30\%$ background). 
The arrival times of these photons were recalculated to the solar system barycenter using 
the SAS {\tt barycen} tool. As with the {\it Chandra} data, we used the $Z_{m}^2$ test 
\citep{buc83} to search for periodicities in the $5\times10^{-5}-80$ Hz frequency range.  
We calculated $Z_{1}^2$ at $3.5~\times~10^{7}$ equally spaced frequencies, which corresponds 
to oversampling by a factor of 10 compared to the expected width $T_{\rm span}^{-1}\approx 
22~\mu$Hz of the $Z_1^2$ peaks, and guarantees that we miss no peaks. The most significant 
peaks we find have $Z_{1}^{2} = 31.82, 31.68,$ and $30.56$ at $f=72.363177(5) {\rm Hz}$, 
$12.011894(5) {\rm Hz}$, and $3.006574(5) {\rm Hz}$. However, even for  the maximum value 
of $Z_{1}^{2}=31.82$, the corresponding significance is low: only 56.5\%, for the number 
of independent trials $\mathcal{N}=f_{\rm max} T_{\rm{span}}\approx 3.5\times 10^{6}$. 
Therefore, most likely these peaks are due to the noise.

We also calculated $Z_{m}^{2}$ for $m=1,2$ around the tentative pulsation frequency ($f = 
0.1336185$--$0.1336999$ Hz) suggested by the {\it Chandra} data \citep{park06}. We find a 
broad peak,  $Z_1^2=10.06$ at $f=0.133661(5) {\rm Hz}$, overlapping with the {\it Chandra} 
peak. However, even a single trial significance (assuming that the periodicity found in 
the {\it Chandra} data is real and hence the pulsation frequency is known) for the peak 
in the pn data is only marginal ($2.7\sigma$ with the corresponding $f_{\rm p}=24\%$), 
while for a blind search significance of this peak is negligible.

Finally, we searched for periodicity in the combined {\it Chandra} and {\it XMM-Newton}
data. Allowing for a non-zero period derivative typical for anomalous X-ray pulsars
(AXPs), we calculate $Z_1^2$ on two-dimensional grid: $f=5\times10^{-5}-0.15$ Hz 
and $\dot{P}=0-3\times10^{-13}$ s s$^{-1}$. The maximum value of $Z_{1}^2$ is 
29.9  at $f = 0.1336615$ Hz and $\dot{P}=5\times10^{-14}$ s s$^{-1}$. Although the 
frequency corresponding to the maximum $Z_1^2$ is consistent with that of the peak found 
in the {\it Chandra} data alone, the significance of the peak found in the joint data is 
very low. Thus, we conclude that the tentative 7.48 s periodicity reported with \
{\it Chandra} is not confirmed by the {\it XMM-Newton} data.

\section{\label{sec:snr} Spectral Analysis of the Supernova Remnant}

{\it XMM-Newton} images of SNR G330.2+1.0 are presented in Fig.~\ref{fig:fig2}. Since 
the small-window mode was used for the pn, the SNR is detected only on the MOS detectors. 
The {\it Chandra} image has revealed that G330.2+1.0 is a shell-type SNR with enhanced 
emission in the thin SW and NE parts of the shell \citep{park06}. Our {\it XMM-Newton} 
images confirm this general morphology. We further reveal spectral variations across the 
SNR: i.e., the E region of the shell is softer than other regions (Fig.~\ref{fig:fig3}). 
Also, there is a faint hard extended feature at $\sim$2$'$ SW from the CCO (marked with
an arrow in Figs.~\ref{fig:fig2} and \ref{fig:fig3}). These features were not clearly 
seen in the {\it Chandra} data because of their positions in the ACIS-I chip gaps.   

We extracted the spectra from bright portions of the SNR shell in SW and NE 
(Fig.~\ref{fig:fig4}). The SW spectrum was extracted from the $\sim$1$'$ $\times$
3$'$ brightest filament in the SW shell, which contains $\sim$1700 counts ($\sim$25\% 
of them are background) for the MOS1+MOS2 data. This SW shell contains $\sim$1300 counts
(including $\sim$13\% background) for the ACIS-I data. The NE spectrum was extracted 
from a circular region ($\sim$30$^{\prime\prime}$ in radius) in the NE parts of the shell. 
This region contains $\sim$570 counts ($\sim$30\% of them are background) and $\sim$540 
counts (including $\sim$30\% background) for the MOS1+MOS2 and the ACIS-I, respectively.
We used the 0.5--10 keV band spectrum for the spectral analysis, and the source spectra 
were binned to contain a minimum of 20 counts per energy bin. Because of the non-uniform
particle background across the MOS detectors, the background spectrum for the {\it XMM-Newton}
data was carefully selected for faint extended sources. We chose a few background 
regions close to the source regions while avoiding any detected (by {\it Chandra}) 
point sources. We find generally consistent results between the background subtracted 
{\it XMM-Newton} spectra and the {\it Chandra} data. Thus, we believe that our background 
estimates for the {\it XMM-Newton} data are acceptable. 

The X-ray spectra of G330.2+1.0 extracted from the SW and NE regions are shown in 
Fig.~\ref{fig:fig5}. Our {\it Chandra} and {\it XMM-Newton} data show featureless 
continuum-dominated spectra for the bright SW and NE filaments. For each region, we 
performed a simultaneous PL model fit for all the three spectra obtained by the MOS1, MOS2, 
and ACIS-I (Fig.~\ref{fig:fig5}). The best-fit parameters are presented in Table~\ref{tbl:tab2}. 
The high absorbing column for the SNR shell is consistent with that for the CCO J1601, 
supporting the SNR-CCO association. The PL photon indices are typical for synchrotron 
emission from highly accelerated relativistic electrons. Thus, we fit these SW and NE shell 
spectra with the SRCUT model, which describes X-ray synchrotron emission from the 
shock-accelerated electrons that are also responsible for the observed radio counterpart 
\citep{rey98,rk99}. We assume the radio spectral index $\alpha$ = 0.3 (where $S_{\nu}$ 
$\propto$ $\nu$$^{-\alpha}$) as measured from the entire SNR \citep{green01}. The results 
of the SRCUT model fits are presented in Table~\ref{tbl:tab3}. 

The spectrally-hard, extended emission feature at $\sim$2$'$ SW of the CCO is faint: we obtain 
$\sim$490 counts from this feature (MOS1+MOS1) in which $\sim$60\% of the photons are the 
background. There is no evidence for line features, and the X-ray spectrum may be fitted 
by a PL of $\Gamma$ $\sim$ 2 with $N_{\rm H}$ fixed at 2.6 $\times$ 10$^{22}$ cm$^{-2}$. 
This overall spectral shape, the filamentary morphology and size (about a few $d_5$ pc), 
and the proximity to J1601 raise an intriguing possibility that this feature might be 
related to the CCO (e.g., the pulsar wind nebula). Alternatively, it could be a part
of the SNR shell. However, reliable spectral modeling of this faint feature is difficult 
because of the poor photon statistics. Thus, we do not attempt any further analysis or 
discussion on this feature. Follow-up deep X-ray observations are required to reveal the 
origin of this potentially intriguing feature. 

On the other hand, the E parts of the SNR shell are spectrally softer than other regions
(Figs.~\ref{fig:fig2} and \ref{fig:fig3}). The E region spectrum is extracted from an 
$\sim$1$\farcm$3 $\times$ 2$'$ region in the E parts of the shell (Figs.~\ref{fig:fig4}
and \ref{fig:fig6}). This region contains $\sim$730 counts ($\sim$45\% of them are background) 
for the MOS1+MOS2 data. Since the central part of this region falls in the ACIS-I chip gap, 
we use only the {\it XMM-Newton} data for the spectral analysis. The best-fit PL photon 
index for region E is significantly steeper ($\Gamma$ $\sim$ 4--5, $\chi^2_{\nu}$ $\sim$ 
1.3--1.4, depending on the assumed $N_{\rm H}$) than those for the SW and NE regions. In fact, 
the PL of $\Gamma$ = 2.3 (as an average for SW and NE shell) cannot fit the observed spectrum 
of region E ($\chi^2_{\nu}$ $\sim$ 1.8--3.2, depending on assumed $N_{\rm H}$) because of a
soft excess emission at $E$ $\la$ 2 keV. This suggests the presence of soft thermal 
emission in the E part of the shell. Thus, we fit the region E spectrum with a plane-shock 
(PSHOCK) model \citep{bor01}. Since the photon statistics are poor for this faint feature, 
we fixed the metal abundances at the solar values \citep{ag89}. Initially we fit the observed 
spectrum with a single PSHOCK model, assuming that X-ray emission in region E is entirely thermal 
in origin. Then, we used a two-component model (PSHOCK + PL) assuming that there is underlying 
nonthermal emission as seen in the SW and NE filaments. Results from these two model fits 
are statistically indistinguishable ($\chi^2_{\nu}$ $\sim$ 1.2 for either model fit). The main 
difference is that the best-fit electron temperature ($kT$ = 1.4$^{+0.9}_{-0.6}$ keV) for the 
single PSHOCK model appears to be somewhat higher than that for the PSHOCK + PL model ($kT$ = 
0.7$^{+1.3}_{-0.3}$ keV). The best-fit volume emission measure ($EM$) for the two-component 
model is higher by a factor of $\sim$2 than that for the single PSHOCK model. Since the 
uncertainties of these measurements are large because of poor photon statistics, it is 
difficult to discriminate these modeled parameters. Thus, we assume plausible ranges of the 
best-fit electron temperature and emission measure in the following discussion, based on the 
these two models: i.e., $kT$ $\sim$ 0.7--1.4 keV and $EM$ $\sim$ 0.6--1.4 $\times$ 10$^{56}$ 
cm$^{-3}$. The results from the PSHOCK + PL model fit for the E region are summarized in 
Table~\ref{tbl:tab2}.

\section{\label{sec:disc} Discussion}

\subsection{\label{subsec:cco} Characteristics of the Central Compact Object}

The previous {\it Chandra} data analysis of J1601 revealed characteristics of 
a thermal spectrum, a location at the center of SNR G330.2+1.0, a pointlike morphology 
without any extended nebulosity, the absence of counterparts at other wavelengths, and 
a large limit on the X-ray-to-optical flux ratio \citep{park06}. J1601 also shows no 
evidence for long-term variability, which is confirmed by our new {\it XMM-Newton} 
data showing a constant X-ray flux ($f_{\rm 1-10 keV}$ $\sim$ 1.2 $\times$ 10$^{-13}$ 
ergs cm$^{-2}$ s$^{-1}$) over the $\sim$2 yr period. Thus, J1601 is most likely the CCO 
associated with SNR G330.2+1.0. Park et al. (2006) noted that the BB temperature is higher 
than the surface temperature expected from the standard cooling of a young neutron star, 
and that the estimated emitting area is too small to be a neutron star. It was also 
noted by Park et al. (2006) that the suggested candidate pulsations with a long period, 
if confirmed, would have been typical for an AXP. 

Our results from the {\it XMM-Newton} and {\it Chandra} data analysis indicate that the hot 
component emission ($T^{\infty}_{\rm h}$ $\sim$ 2.5--5.5 MK, depending on the assumed $B$) 
must originate from a small region of $R_{\rm h}$ $\sim$ 0.4--2 $d_5$ km. The estimated 
size of the hot region varies depending on the assumed values of the magnetic field and the 
distance to the CCO. Nonetheless, within the ranges of parameters that we consider in this 
work ($B$ = 0, 10$^{12}$, or 10$^{13}$ G, and $d$ = 5--10 kpc), the size of the hot region is 
significantly smaller than the canonical size of the neutron star; i.e., the largest area could 
be $R_{\rm h}$ $\sim$ 4 km, where $B$ = 0 and $d$ = 10 kpc. A small hot region(s) has been 
suggested in other CCOs, probably indicating X-ray emission from a locally-heated region such 
as the hot polar cap (e.g., Pavlov et al. 2000). On the other hand, the estimated surface 
temperature of the neutron star is significantly lower ($T^{\infty}_{\rm s}$ $<$ 1.5 MK) than 
that of the hot region. According to the standard cooling curves of a neuron star (e.g., Tsuruta 
1998; Yakovlev \& Pethick 2004), this temperature limit corresponds to a lower limit of several 
10$^{2}$--10$^4$ yr for the neutron star's age. This neutron star age is in plausible 
agreement with the estimated age of SNR G330.2+1.0 (see \S\S~\ref{subsec:nonthermal} and
\ref{subsec:thermal}, and Torii et al. 2006). The overall characteristics such as the low 
$T^{\infty}_{\rm s}$, the high $T^{\infty}_{\rm h}$, and the small $R_{\rm h}$ are consistent 
with those found in the prototype CCOs in Galactic SNRs such as Cas A and Vela Jr. 
\citep{pav00,pav01}. 

Since the X-ray flux from the small hot region contributes a significant fraction of the
observed flux ($>$ 50\% of the total flux in the 1--10 keV band), the observed X-ray 
emission from J1601 may be expected to pulsate. However, our {\it XMM-Newton} data do not 
show any conclusive evidence for pulsations, indicating that the previously suggested 
pulsations are unlikely real. We note that the low photon statistics in the EPIC-pn 
data are not sufficient to detect pulsations with an intrinsic pulsed-fraction $f_{\rm p}$ 
$\la$40\%.  With the combined {\it Chandra} and {\it XMM-Newton} data, the detection of 
pulsations with $f_{\rm p}$ $\la$25\% is not feasible. Thus, the presence of an X-ray pulsar 
for J1601 is not ruled out by the current data. The neutron star's magnetic field, which 
would provide critical information on the nature of the object, remains unknown. Deep X-ray 
observations of J1601 are required to make conclusive remarks on the nature of J1601 such 
as pulsations, magnetic field, age, and the origin of its X-ray emission. 

\subsection{\label{subsec:nonthermal} Nonthermal X-Ray Emission of the Supernova Remnant}

Our joint spectral analysis of the {\it XMM-Newton} and {\it Chandra} data of G330.2+1.0
shows that X-ray emission from the bright filaments of the SNR shell is dominated
by a PL continuum. We find that this PL spectrum prevails for the most parts of the SNR, 
which was also suggested by a previous study \citep{torii06}. The best-fit PL model for the 
bright SW and NE regions of the shell indicates photon indices of $\Gamma$ $\sim$ 2.1--2.5 
which are typical for synchrotron emission from shock-accelerated relativistic electrons. 
Although thermal plasma models may also fit the observed spectra, the estimated electron 
temperatures are high ($kT$ $\sim$ 4--5 keV), and low metal abundances ($\la$ 0.1 solar, 
Anders \& Grevesse 1989) are required. While a thermal origin of X-ray emission from the 
SNR shell may not be completely ruled out by the current data, the estimated plasma 
temperature and abundances appear to be unusual for SNRs. Thus, except for region E 
(\S~\ref{subsec:thermal}), we discuss this SNR based on the nonthermal interpretations of 
X-ray emission. 

According to our SRCUT model fits of the SW and NE filaments, the best-fit exponential roll-off 
frequency, $\nu_{\rm rolloff}$ $\sim$ 1.6--3.3 $\times$ 10$^{17}$ Hz, is relatively high among 
Galactic SNRs \citep{rk99}, while being similar to those for SN 1006 \citep{bamba03} and the 
bright TeV $\gamma$-ray emitting SNR G347.3--0.5 \citep{laz04}. If the particle (electron) 
acceleration is limited by synchrotron losses, the cutoff frequency corresponding to the 
maximum electron energy $E_{\rm max}$ is $\nu_{\rm m}$(loss) $\propto$ $B$ $E^2_{\rm max}$(loss). 
Since $E_{\rm max}$(loss) $\propto$ $B^{-{1\over2}}$, $\nu_{\rm m}$(loss) is independent of $B$, 
and depends only on the shock velocity: e.g., assuming a strong shock of the compression ratio 
of $>$ 4 and the shock normal perpendicular to $B$, the cutoff frequency is $\nu_{\rm m}$(loss) 
$\ga$ 3 $\times$ 10$^{16}$ $\eta$ $v^2_3$ Hz, where $v_3$ is the shock velocity in units of 
10$^3$ km s$^{-1}$, and the ratio of the electron scattering mean free path to the gyroradius 
$\eta$ $\geq$ 1 (e.g., Reynolds 1998; Lazendic et al. 2004). As discussed below and in
\S~\ref{subsec:thermal}, the shock velocity appears to be roughly $v_s$ $\sim$ 4000 km s$^{-1}$ 
for G330.2+1.0, and thus we estimate $\nu_{\rm m}$(loss) $\ga$ 5 $\times$ 10$^{17}$ Hz. 
Unless the shock velocity is much higher and/or the particle acceleration is inefficient 
($\eta$ $\gg$ 1), the estimated $\nu_{\rm m}$(loss) is comparable with the observed 
$\nu_{\rm rolloff}$, suggesting that the particle acceleration of electrons in G330.2+1.0 is 
likely limited by synchrotron losses rather than the age of the SNR. The peak frequency of a 
synchrotron emitting electron is $\nu_{\rm p}$ = 1.8 $\times$ 10$^{18}$ $E^2_{\rm e} B$ Hz, 
where $B$ is the postshock magnetic field perpendicular to the shock normal, and $E_{\rm e}$ 
is the electron energy. For $\nu_{\rm p}$ $\sim$ 7 $\times$ 10$^{17}$ Hz (or $\sim$3 keV) 
representing the typical X-ray photons based on the observed spectrum of the nonthermal 
filaments in G330.2+1.0, the corresponding electron energy is $E_{\rm e}$ = 0.62 $B^{-{1\over2}}$ 
ergs. The characteristic synchrotron loss time scale for such electrons can then be estimated 
to be $\tau_{\rm loss}$ = 630 $E^{-1}_{\rm e}$ $B^{-2}$ s = 1017 $B^{-{3\over2}}$ s. 

We estimate $\tau_{\rm loss}$ by measuring the widths of the bright nonthermal filaments 
of G330.2+1.0 using {\it Chandra} images (Fig.~\ref{fig:fig7}). We construct projected 
intensity profiles across the bright SW filaments by averaging the photon counts (in 
4$^{\prime\prime}$ pixel bins) over the 40$^{\prime\prime}$ segments along the filaments. 
We fit these 1-D intensity profiles with a Gaussian to estimate the widths of the filaments. 
We note that high resolution {\it Chandra} images of bright X-ray synchrotron filaments
in young SNRs show typical substructures of a broad exponential downstream region and 
a much steeper flux decay in the upstream (e.g., Bamba et al. 2003). G330.2+1.0 is more 
distant than other young SNRs (that show bright nonthermal filaments), and the X-ray shell is 
relatively faint, which does not allow us to resolve such a substructure. Since the 
downstream region is observed to dominate the width of the filaments, we assume a negligible 
contribution from the upstream emission in the widths of the filaments to measure the 
{\it downstream} widths of the filaments with a simple Gaussian model. The measured widths are 
$\sim$12$^{\prime\prime}$--16$^{\prime\prime}$ (FWHM) which correspond to physical sizes 
$D$ $\sim$ 0.3--0.4 $d_5$ pc. Because of the far distance and faint surface brightness
of G330.2+1.0, our width measurements could be an overestimate from superpositions of 
thinner filaments. Nonetheless, the estimated widths are comparable with an average value 
for the individual filaments in SN 1006 ($\sim$0.2 pc, Bamba et al. 2003). Therefore, 
we take our measurements as a first-order estimate, and certainly as an upper limit.

The advection distance of the downstream electrons from the shock is $D_{\rm ad}$ = $v_s$ 
$\tau_{\rm loss}$ $r^{-1}$, where $r$ is the compression ratio in the shock. Since the 
direct measurements of the shock velocity of G330.2+1.0 are not available, we consider 
some plausible estimates for the shock velocity based on several independent approaches. 
Assuming an electron-ion temperature equipartition in the postshock region, the detected 
thermal emission of G330.2+1.0 (region E) implies $v_s$ $\sim$ 1000 km s$^{-1}$ 
(\S~\ref{subsec:thermal}).  This value may be considered as a lower limit for $v_s$, 
because the assumed temperature equilibration between electrons and ions may have not been 
established in relatively young SNRs with $v_s$ $\ga$ several 10$^2$ km s$^{-1}$ \citep{ghav07}. 
G330.2+1.0 shows similar characteristics (e.g., the SNR age, $\nu_{\rm rolloff}$, and the 
physical width of the nonthermal filaments etc.) to those of G347.3--0.5 and SN 1006 in 
which the shock velocities are high ($v_s$ $\sim$ 3000--4000 km s$^{-1}$, e.g., Parizot et al. 
2006 and references therein). The ambient density for G330.2+1.0 ($n_0$ $\sim$ 0.1 cm$^{-3}$, 
\S~\ref{subsec:thermal}) is not unusually high compared with other SNRs (e.g., Bamba 
et al. 2003). Thus, the actual shock velocity of G330.2+1.0 is likely higher than $v_s$ 
$\sim$ 1000 km s$^{-1}$, perhaps close to $v_s$ $\sim$ 3000--4000 km s$^{-1}$. In fact, 
models predict high shock velocities of $v_s$ $\ga$ 2000 km s$^{-1}$ for an efficient 
particle acceleration (e.g., Ellison et al. 2000;2004). Although a small sample is used, 
an empirical relationship between $\nu_{\rm rolloff}$ and the physical width of the 
nonthermal filaments $D$ is derived to be $\nu_{\rm rolloff}$ $D^{-2}$ = 2.6 $\times$ 
10$^{27}$ $\tau^{-2.96}_{\rm SNR}$ for several young historical SNRs \citep{bamba05}. 
This empirical relation suggests an SNR age $\tau_{\rm SNR}$ $\sim$ 1000--1200 yr for 
G330.2+1.0 for the measured $\nu_{\rm rolloff}$ $\sim$ 2--3 $\times$ 10$^{17}$ Hz. 
The inferred young age and the low ambient density suggest that the SNR may be in a 
free-expansion or an adiabatic phase, or could be in transition between the two. Assuming 
an adiabatic phase, the suggested SNR ages imply $v_s$ $\sim$ 2300--2800 km s$^{-1}$ for 
the SNR radius of $R$ = 7.3 pc (see \S~\ref{subsec:thermal} for the SNR radius). For 
a free-expansion phase, $v_s$ $\sim$ 5800--7000 km s$^{-1}$ is implied. These high 
velocities are consistent with those estimated for young SNRs showing an efficient particle 
acceleration (e.g. Parizot et al. 2006 and references therein). Thus, as a rough estimate 
by {\it averaging} several values discussed above, we for simplicity adopt a shock 
velocity $v_s$ $\sim$ 4000 km s$^{-1}$ for G330.2+1.0. This shock velocity is admittedly 
not a measurement and thus only a crude first-order estimate. (We would allow a factor 
of $\sim$2 uncertainty in this velocity estimate, and within this range, our conclusions 
as discussed below are not affected.) 

Assuming $r$ $\sim$ 5--8 for an efficient particle acceleration (e.g., Ellison et al. 2007), 
we estimate $\tau_{\rm loss}$ $\sim$ 350--600 yr for the measured $D$ $\sim$ $D_{\rm ad}$ $\sim$ 
0.3 $d_5$ pc, and thus $B$ $\sim$ 14--20 $\mu$G. The maximum electron energy can be estimated 
by $E_{\rm max}$ = 2.5 $\times$ 10$^{-7}$ $\nu_{\rm rolloff}^{1\over2}$ $B^{-{1\over2}}$ 
TeV = 100--144 $B^{-{1\over2}}_{{\mu}{\rm G}}$ TeV, where $B_{{\mu}{\rm G}}$ is the 
postshock magnetic field in units of $\mu$G \citep{rk99,laz04}. Thus, $E_{\rm max}$ $\sim$ 
22--38 TeV (depending on measured $\nu_{\rm rolloff}$) is derived. In addition, if we 
consider a geometrical projection effect in measuring the widths of the nonthermal filaments 
(e.g., the observed width is $\sim$ 4.6 $\times$ the actual width assuming a spherical shock 
with an exponential emission profile, Ballet 2006), the estimated $B$ can be a few times 
higher ($\sim$50 $\mu$G).

The estimated $E_{\rm max}$ for G330.2+1.0 suggests that this SNR is a candidate 
$\gamma$-ray source. For instance, the $\gamma$-ray emission by the inverse Compton 
(IC) scattering off interstellar photons can be estimated by $E_{\gamma}$ $\sim$ 5.1 
$\times$ 10$^{-12}$ $E_{\star}$ $E^2_{\rm e}$ eV, where $E_{\gamma}$ is the average final 
energy of the up-scattered photons, and $E_{\star}$ is the typical energy for the seed 
photons \citep{tat08}. Using $E_{\star}$ $\sim$ 7 $\times$ 10$^{-4}$ eV for the cosmic 
microwave background (CMB) and $E_{\rm e}$ = $E_{\rm max}$ $\sim$ 30 TeV, we estimate 
$E_{\gamma}$ $\sim$ 3 TeV. However, G330.2+1.0 is not identified in the H.E.S.S. 
Galactic plane survey catalog \citep{aha06a}. It is probably because G330.2+1.0 is
more distant and thus apparently fainter than other TeV-bright SNRs (e.g.,
G347.3--0.5 and G266.2--1.2). The IC to synchrotron flux ratio $f_{\rm IC}$/$f_{\rm syn}$ 
= 8$\pi$$U_{\rm rad}$/$B^2$ $\sim$ 10 $B^{-2}_{{\mu}{\rm G}}$ $\sim$ 0.004--0.1 (where the 
energy density of the seed CMB photons $U_{\rm rad}$ $\sim$ 0.25 eV cm$^{-3}$) for the 
plausible range of $B$ $\sim$ 10--50 $\mu$G in G330.2+1.0. These $f_{\rm IC}$/$f_{\rm syn}$ 
are in fact similar to the observed $f_{\rm TeV}$/$f_{\rm X}$ for SNRs G347.3--0.5 and 
G266.2--1.2 (e.g., Matsumoto et al. 2007 and references therein). Then, the overall 
X-ray flux of $f_{\rm syn}$ $\sim$ 10$^{-11}$ ergs cm$^{-2}$ s$^{-1}$ for G330.2+1.0 
\citep{torii06} implies $f_{\rm IC}$ $\sim$ 10$^{-13}$--10$^{-12}$ ergs cm$^{-2}$ s$^{-1}$.
The sky position of G330.2+1.0 was at the edge of the H.E.S.S. survey, in which the 
exposure was short ($<$5 hr). Considering the small angular size ($\sim$10$'$) of 
G330.2+1.0, which is close to the point spread function of the H.E.S.S. (several arcminutes), 
and the short exposure in the survey, the estimated IC flux is likely close to or below 
the H.E.S.S. detection limit of $f$ $\sim$ 10$^{-12}$ ergs cm$^{-2}$ s$^{-1}$ at $E$ $\ga$ 
1 TeV (e.g., Aharonian et al. 2005). Thus, if the $\gamma$-ray emission from G330.2+1.0 is 
dominated by the IC process of the same electrons to produce X-ray synchrotron emission, 
the non-detection of G330.2+1.0 with the current H.E.S.S. survey data is not surprising. 
A deep search of $\gamma$-ray emission for G330.2+1.0 using ground-based TeV telescopes 
and {\it Fermi} (formerly {\it GLAST}) is warranted.
  
It is notable that nonthermal X-ray emission in G330.2+1.0 is generally anti-correlated
with the radio emission \citep{torii06}. Our high resolution {\it Chandra} and {\it 
XMM-Newton} images reveal that there actually exist radio counterparts for the bright X-ray 
filaments in SW and NE, but the radio emission is faint (Fig.~\ref{fig:fig3}). The brightest 
radio emission is in the E parts of the SNR, where X-ray emission is faint and spectrally 
soft (Fig.~\ref{fig:fig3}). Thus, the bright radio emission likely traces high density 
regions where soft (thermal) X-ray emission is enhanced. Based on our SRCUT model fits, 
X-ray emission in SW and NE filaments implies the 1 GHz radio flux of 0.7--1.5 $\times$ 
10$^{-4}$ Jy, while the MOST 843 MHz image of the SNR suggests $\sim$0.1 Jy for these regions 
(assuming that the total 1 GHz flux for the entire SNR is 5 Jy, Green 2001). Although our 
radio flux estimates are crude and should be considered only as an order-of-magnitude
approximation based on a simple ``normalization'' of the total image intensity to the 
area corresponding to the X-ray-bright SW and NE filaments, the discrepancy is substantial 
by three orders of magnitudes, and should thus be real. We do not have an immediate answer 
as to what causes the large difference between the modeled and observed radio fluxes 
corresponding to the X-ray bright filaments. One speculation is that the radio spectral 
index might not be uniform across the SNR. While the overall radio spectrum is fitted 
by $\alpha$ = 0.3, the faint radio filaments corresponding to the bright X-ray shell 
might have a steeper spectrum. For instance, if we assume a plausible range of the 
{\it observed} radio flux $\sim$0.01--0.1 Jy for the SW region and vary the radio spectral 
index in our SRCUT model fit, we obtain a best-fit $\alpha$ $\sim$ 0.53--0.66 
($\chi^2_{\nu}$ = 1.2). These radio spectral indices are not unusual for shell-type SNRs 
\citep{green01}. The best-fit roll-off frequencies are high, but are poorly constrained 
($\nu_{\rm rolloff}$ = 13$^{+93}_{-9}$ $\times$ 10$^{17}$ Hz when the 1 GHz radio flux 
of 0.1 Jy is assumed, and $\nu_{\rm rolloff}$ = 8$^{+25}_{-6}$ $\times$ 10$^{17}$ Hz 
for the radio flux of 0.01 Jy). Although the high roll-off frequency, $\nu_{\rm rolloff}$ 
$\sim$ 10$^{18}$ Hz, implies somewhat higher estimates for the shock velocity and the 
maximum electron energy, these changes do not make a significant effect on our conclusions 
presented here. High resolution radio observations with a deep exposure would be essential 
to study the detailed relationship between the X-ray and the radio emission in this SNR. 

\subsection{\label{subsec:thermal} Thermal X-Ray Emission of the Supernova Remnant}

In the E region, soft thermal emission is a significant component in the observed X-ray 
spectrum. The best-fit electron temperature is $kT$ $\sim$ 0.7--1.4 keV, depending on 
models (\S~\ref{sec:snr}). The best-fit ionization timescale appears to be high ($n_{\rm e}t$ 
$\ga$ 10$^{13}$ cm$^{-3}$ s) suggesting that the plasma could be in collisional ionization 
equilibrium, but the $n_{\rm e}t$ parameter is not well-constrained because of the low photon 
statistics. Detecting thermal emission in SNRs in which nonthermal emission dominates is 
critical to reveal the environmental conditions (e.g., ambient density) and the supernova 
energetics that should have affected the SNR evolution and the particle acceleration. In fact, 
G330.2+1.0 is the only example to reveal thermal X-ray emission among the four Galactic SNRs 
which have been known to be dominated entirely by nonthermal X-rays (see \S~{\ref{sec:intro}). 
Therefore, although it is difficult to perform a thorough spectral analysis of thermal 
emission and to draw firm conclusions on the nature of the SNR because of the poor photon 
statistics for the faint thermal component, we present a brief discussion on some fundamental 
SNR parameters based on our spectral analysis of region E.

Based on the best-fit volume emission measure ($EM$ = $n_{\rm e}$$n_{\rm H}$$V$, where 
$n_{\rm e}$, $n_{\rm H}$, and $V$ are the postshock electron, proton densities, and the
X-ray emitting volume, respectively), we estimate $n_{\rm e}$ $\sim$ 0.4--0.5 
$f^{-{1\over2}}$$d^{-{1\over2}}_5$ cm$^{-3}$ (where $f$ is the X-ray emitting volume
filling factor). These postshock electron densities correspond to the preshock hydrogen 
density $n_{\rm 0}$ $\sim$ 0.1 $f^{-{1\over2}}$$d^{-{1\over2}}_5$ cm$^{-3}$. In these 
estimates, we assume $n_{\rm e}$ = 1.2 $n_{\rm H}$ for the mean charge state with normal 
composition, and $n_{\rm H}$ = 4$n_0$ for a strong shock. We use the emission volume 
$V$ $\sim$ 4 $\times$ 10$^{56}$ cm$^{-3}$ assuming that the path-length through region E 
is comparable to the physical size corresponding to the angular size ($\sim$2$'$) of region 
E at $d$ = 5 kpc. Assuming an ion-electron temperature equilibration, the measured electron 
temperature implies a shock velocity of $v_s$ $\sim$ 800 ($kT$ = 0.7 keV) -- 1100 ($kT$ = 
1.4 keV) km s$^{-1}$. However, equipartition of the electron-ion temperatures may not 
have been reached, and thus the actual shock velocity could be higher than $v_s$ $\sim$ 1000 
km s$^{-1}$, probably by a factor of a few (\S~\ref{subsec:nonthermal}). We estimate the SNR 
radius of $R$ $\sim$ 5$'$ (the half of the angular distance between the the bright SW and NE 
filaments), which corresponds to the physical distance of $\sim$7.3 $d_5$ pc. Then, assuming 
an adiabatic phase for the SNR, we apply the Sedov solution to derive the SNR age $\tau_{\rm 
SNR}$ $\sim$ 1100 $d_5$ yr (e.g., for $v_s$ $\sim$ 2500 km s$^{-1}$, \S~\ref{subsec:nonthermal}). 
For a free-expansion phase, the SNR age is also derived to be $\tau_{\rm SNR}$ $\sim$ 1100 yr 
(e.g., for $v_s$ $\sim$ 6500 km s$^{-1}$, \S~\ref{subsec:nonthermal}). Using a Sedov solution, 
the explosion energy is estimated to be $E_0$ $\sim$ 2--9 $\times$ 10$^{50}$ $d^{5\over2}_5$ 
ergs for $\tau_{\rm SNR}$ $\sim$ 1000--2000 yr. 

\section{\label{sec:sum} Summary and Conclusions}

Based on the {\it ASCA} data, the overall X-ray emission from SNR G330.2+1.0 was suggested
to be continuum-dominated with no evidence for line features \citep{torii06}. 
The high resolution {\it Chandra} images subsequently revealed that X-ray emission from this
SNR originates primarily from the thin shell with enhanced filaments in the SW and 
NE parts of the shell \citep{park06}. Park et al. (2006) have also discovered the
CCO J1601 at the center of the SNR. We performed follow-up observations of G330.2+1.0 with
{\it XMM-Newton} to investigate the nature of the CCO and the SNR. Although 
our spectral and temporal analyses of J1601 and G330.2+1.0 are limited by poor photon 
statistics of the {\it XMM-Newton} data caused by significant contamination from flaring 
particle background, we find several important characteristics of these objects utilizing
the {\it XMM-Newton} and {\it Chandra} data.

The X-ray spectrum of J1601 can be described by two-component neutron star atmosphere 
models. X-ray emission primarily originates from a small hot region ($R$ $\sim$ 0.4--2 km, 
$T$ $\sim$ 2.5--5.5 MK). The rest of the neutron star's surface is cooler ($R$ $\sim$ 10 km, 
$T$ $<$ 1.5 MK), suggesting an $\ga$10$^{3-4}$ yr old neutron star based on the standard cooling
models. The neutron star atmosphere models do not provide useful constraints on the magnetic 
field of J1601 with the current data. The previously suggested pulsations ($P$ $\sim$ 7.48 s) 
are not confirmed by the {\it XMM-Newton} data. These characteristics are similar to those 
found for CCOs in other Galactic SNRs such as Cas A and Vela Jr. The spectrally hard,
faint nebulosity at $\sim$2$'$ SW from the CCO could be the associated PWN, but its true
nature is uncertain with the current data because of the poor photon statistics. 
Follow-up deep X-ray observations are required to reveal the detailed nature of J1601.

Assuming that X-ray emission in the shell of G330.2+1.0 is synchrotron radiation from
the shock accelerated electrons, the roll-off frequency of $\nu_{\rm rolloff}$ $\sim$
1.6--3.3 $\times$ 10$^{17}$ Hz is estimated. It is difficult to measure the shock velocity
with the currently available data. Based on several independent approaches, we make a
rough estimate of the shock velocity $v_s$ $\sim$ 4000 km s$^{-1}$ (with a factor of $\sim$2
uncertainty). Based on this shock velocity and the measured roll-off frequency, we find 
that the particle (electron) acceleration in G330.2+1.0 is likely limited by synchrotron 
losses rather than the SNR age. Using the {\it Chandra} images, we measure the widths of 
the bright nonthermal X-ray filaments ($D$ $\sim$ 0.3--0.4 pc). Using these widths and the 
shock velocity, we estimate the synchrotron loss time of $\tau_{\rm loss}$ $\sim$ 350--600 
yr and the magnetic field of $B$ $\sim$ 10--50 $\mu$G. The maximum electron energy is 
derived to be $E_{\rm max}$ $\sim$ 22--38 TeV. These electron energies suggest that 
G330.2+1.0 is a candidate $\gamma$-ray source (up to $\sim$TeV) by the IC scattering of 
the CMB photons. The non-detection of G330.2+1.0 in the current H.E.S.S. survey with a 
short exposure is perhaps expected, because G330.2+1.0 is more distant and likely 
a fainter $\gamma$-ray source than the bright TeV SNRs like G347.3--0.5 and Vela Jr. 

G330.2+1.0 is particularly intriguing because this is the only SNR in which we detect a 
thermal component among the four Galactic SNRs known to be dominated by nonthermal X-ray 
emission. Although the uncertainties are large due to the poor photon statistics, the 
estimated density ($n_0$ $\sim$ 0.1 cm$^{-3}$) is low, suggesting that $\gamma$-ray emission, 
if it exists, would be dominated by the IC process. The detection of $\gamma$-ray emission as 
well as thermal X-ray emission with high photon statistics from G330.2+1.0 will be essential 
to test and constrain models for $\gamma$-ray production from shock-accelerated particles. 
Follow-up deep observations with X-ray detectors on board {\it XMM-Newton} and {\it Suzaku} 
are necessary for a thorough study of thermal X-ray emission. Deep $\gamma$-ray observations 
using {\it Fermi} and the ground-based TeV telescopes will be critical to reveal the nature 
of nonthermal radiation produced by shock accelerated particles. High resolution radio and 
X-ray observations of G330.2+1.0 with a deep exposure are essential to reveal the origin 
of the apparent inconsistency between the radio and nonthermal X-ray emission, such as the 
radio spectral index variation across the SNR.

\acknowledgments

The authors thank V. E. Zavlin for the helpful discussion on the hydrogen neutron star 
atmosphere models. This work was supported in parts by NASA grant NNX08AW88G and SAO 
grant SV4-74018. POS acknowledges partial support from NASA contract NAS8-03060.
KM was partially supported by the Grant in-Aid for Young Scientists (B) of the MEXT (No. 
18740108). This work makes use of the {\it Supernova Remnant Catalog} by the MOST which is 
operated by the University of Sydney with support from the Australian Research 
Council and the Science Foundation for Physics within the University of Sydney.

\clearpage

\begin{deluxetable}{cccccccc}
\tabletypesize{\footnotesize}
\tablecaption{Best-Fit Parameters of J1601 from two-component neutron star atmosphere models.
\label{tbl:tab1}}
\tablewidth{0pt}
\tablehead{\colhead{$B$} & \colhead{$N_H$} & \colhead{$T^{\infty}_{\rm s}$\tablenotemark{a}} & 
\colhead{$T^{\infty}_{\rm h}$\tablenotemark{a}} & \colhead{$R_{\rm h}$\tablenotemark{b}} & 
\colhead{$f_{1-10~{\rm keV}}$\tablenotemark{c}} & \colhead{$L_{1-10~{\rm keV}}$} & \\
\colhead{(10$^{12}$ G)} & \colhead{(10$^{22}$ cm$^{-2}$)} & \colhead{(10$^6$ K)} &
\colhead{(10$^6$ K)} & \colhead{($d_5$ km)} & \colhead{(10$^{-13}$ ergs cm$^{-2}$ s$^{-1}$)} &
\colhead{(10$^{33}$ $d^2_5$ ergs s$^{-1}$)} & \colhead{$\chi^2$/$\nu$} }
\startdata
0 & 3.15$^{+0.63}_{-0.35}$ & $<$ 1.4 & 2.5$^{+0.3}_{-0.2}$ & 2.1$^{+2.3}_{-0.6}$ & 
1.23$\pm$0.06 & 1.5 & 80.5/78 \\
1 & 3.40$^{+0.52}_{-0.20}$ & $<$ 1.5 & 5.5$^{+0.4}_{-0.3}$ & 0.4$^{+0.1}_{-0.1}$ & 
1.21$\pm$0.06 & 1.9 & 79.9/78 \\
10 & 3.29$^{+0.53}_{-0.88}$ & $<$ 1.5 & 3.7$^{+0.6}_{-0.4}$ & 0.9$^{+0.6}_{-0.3}$ & 
1.21$\pm$0.06 & 1.8 & 79.7/78 \\
\enddata
\vspace{-2mm}
\tablecomments{Errors are at 90\% confidence. $M_{\rm ns}$ and $R_{\rm ns}$ are fixed
at 1.4 $M_{\odot}$ and 10 km, respectively.} 
\tablenotetext{a}{$T^{\infty}_{\rm s}$ and $T^{\infty}_{\rm h}$ are effective temperatures
of the cool neutron star surface and the hot small region, respectively, as measured by a distant 
observer where $T^{\infty}$ = $g_{\rm r}T$ and $g_{\rm r}$ = (1--2$GM_{\rm ns}/{R_{\rm 
ns}}c^2$)$^{1\over2}$.}
\tablenotetext{b}{The radius of the hot region scaled by $d$ = 5 kpc.}
\tablenotetext{c}{The observed flux in the 1--10 keV band. Assuming a Poisson distribution
of the observed photon statistics, 2$\sigma$ statistical errors are quoted. }
\end{deluxetable}

\begin{deluxetable}{ccccccc}
\tabletypesize{\footnotesize}
\tablecaption{Best-Fit Power Law Model Parameters for SNR G330.2+1.0.
\label{tbl:tab2}}
\tablewidth{0pt}
\tablehead{ & \colhead{$N_{\rm H}$} & \colhead{$\Gamma$} & \colhead{$kT$} & \colhead{$n_{\rm e}$t} 
& \colhead{$EM$} & \\
\colhead{Region} & \colhead{(10$^{22}$ cm$^{-2}$)} & & \colhead{(keV)} & \colhead{(10$^{11}$ cm$^{-3}$ s)}
& \colhead{(10$^{56}$ cm$^{-3}$)} & \colhead{$\chi^2$/$\nu$}}
\startdata
SW & 2.60$^{+0.40}_{-0.34}$ & 2.13$^{+0.24}_{-0.22}$ & - & - & - & 163.9/137 \\
NE & 3.04$^{+0.65}_{-0.81}$ & 2.52$^{+0.40}_{-0.54}$ & - & - & - & 51.9/51 \\
E & 2.45$^{+0.72}_{-0.57}$ & 2.3 & 0.70$^{+1.34}_{-0.32}$ & $>$ 5 & 1.4$^{+6.8}_{-1.0}$ &
35.3/30 \\
\enddata
\vspace{-2mm}
\tablecomments{Errors are at 90\% confidence. For region E, parameters from the best-fit 
two component model (plane-shock + power law, where $\Gamma$ = 2.3 is fixed) are presented. 
} 
\end{deluxetable}

\begin{deluxetable}{ccccc}
\tabletypesize{\footnotesize}
\tablecaption{Best-Fit SRCUT Model Parameters for SNR G330.2+1.0.
\label{tbl:tab3}}
\tablewidth{0pt}
\tablehead{ & \colhead{$N_{\rm H}$} & \colhead{${\nu}_{\rm rolloff}$} & \colhead{1 GHz Flux} & \\
\colhead{Region} & \colhead{(10$^{22}$ cm$^{-2}$)} & \colhead{(10$^{17}$ Hz)} & \colhead{(mJy)} & 
\colhead{$\chi^2$/$\nu$}}
\startdata
SW & 2.41$^{+0.17}_{-0.15}$ & 3.3$^{+4.1}_{-1.7}$ & 0.15$\pm$0.01 & 163.7/137 \\
NE & 2.59$^{+0.32}_{-0.28}$ & 1.6$^{+4.0}_{-1.1}$ & 0.07$\pm$0.01 & 51.9/51 \\
\enddata
\vspace{-2mm}
\tablecomments{Errors are at 90\% confidence. The radio spectral index is fixed at 
$\alpha$ = 0.3 (where $S_{\nu}$ $\propto$ $\nu$$^{-\alpha}$).}

\end{deluxetable}


\begin{figure}[]
\figurenum{1}
\centerline{\includegraphics[angle=-90,width=0.7\textwidth]{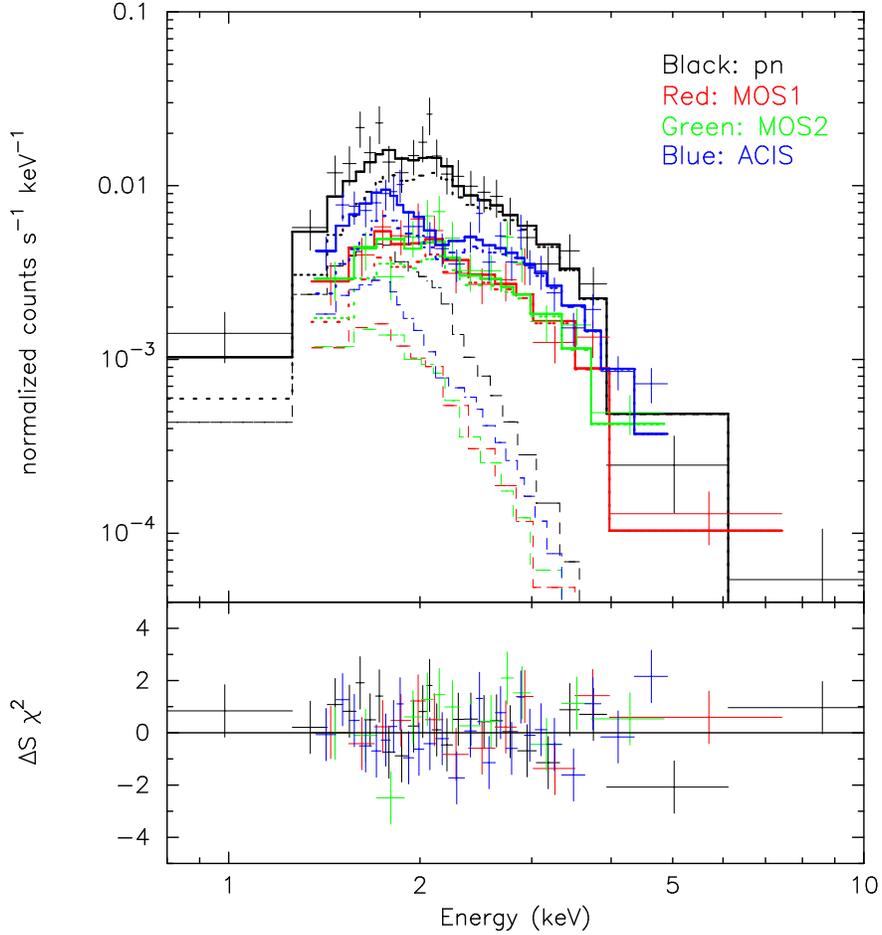}}
\figcaption[]{The X-ray spectrum of J1601 as observed by {\it XMM-Newton} and {\it Chandra}. 
The best-fit two component NSA model with $B$ = 10$^{13}$ G is overlaid. The solid lines
are the best-fit model. The dotted- and dashed-lines are the small hot region and the large 
cool surface components, respectively. The lower panel is the residuals from the best-fit model.
\label{fig:fig1}}
\end{figure}

\begin{figure}[]
\figurenum{2}
\centerline{\includegraphics[angle=0,width=\textwidth]{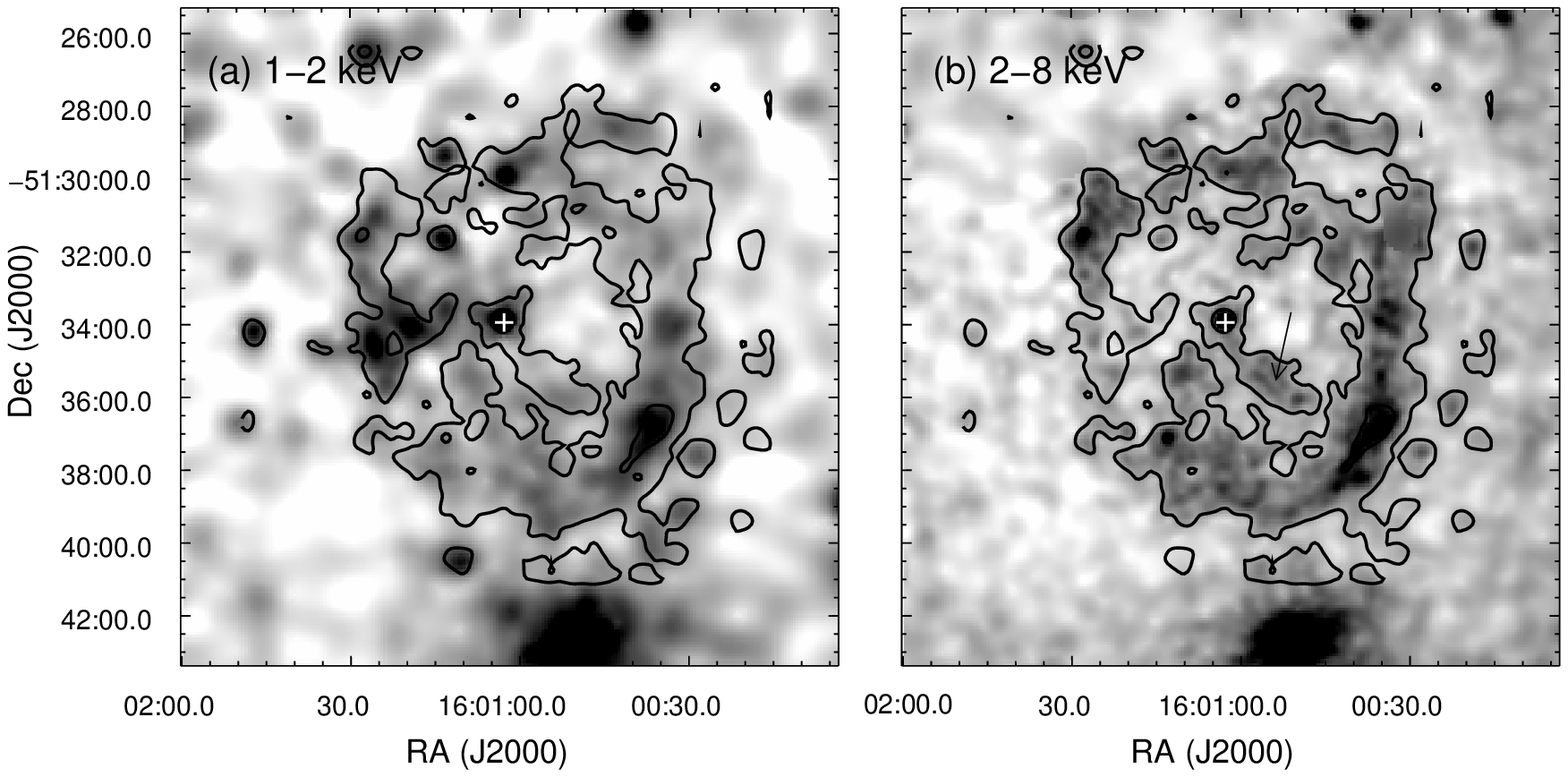}}
\figcaption[]{{\it XMM-Newton} images (MOS1+MOS2) of G330.2+1.0: (a) the soft 
band (1--2 keV), and (b) the hard band (2--8 keV). In (a), the 1.4--1.6 and 1.7--1.8 keV 
bands are excluded to remove the bright instrumental lines at $E$ $\sim$ 1.5 (Al K) and 1.74 
(Si K) keV. Images are exposure-corrected, and darker grey-scales correspond to higher 
intensities. For the purposes of display, the images have been binned into $\sim$5$^{\prime\prime}$ 
pixels, and then adaptively smoothed to achieve a minimum S/N = 7. J1601 is marked with 
a cross at the center of the SNR. Image contours of the broadband (1--8 keV) image 
are overlaid in each panel. In (b), the hard feature seen in Fig.~\ref{fig:fig3} is
marked with an arrow.
\label{fig:fig2}}
\end{figure}

\begin{figure}[]
\figurenum{3}
\centerline{\includegraphics[angle=0,width=0.80\textwidth]{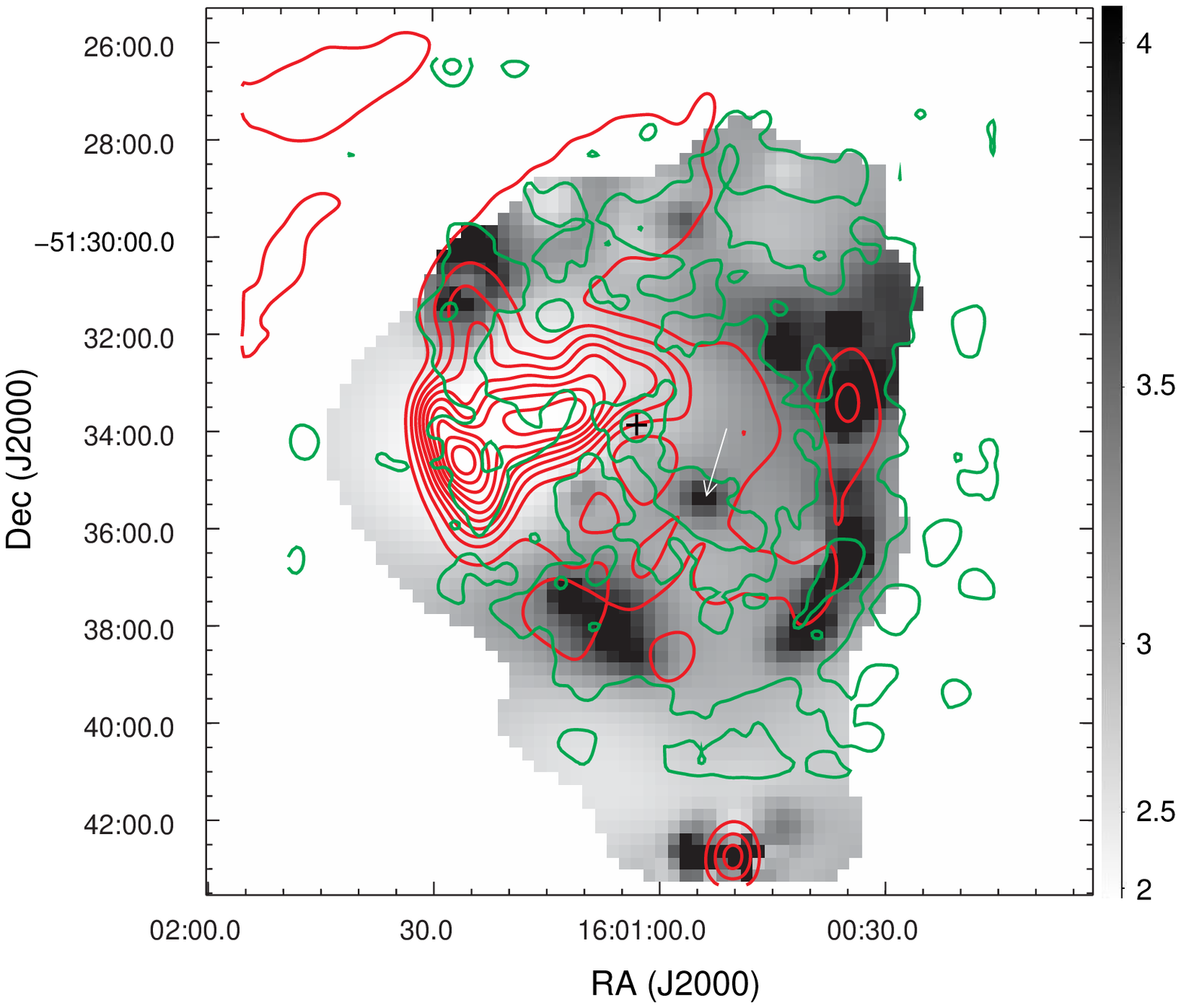}}
\figcaption[]{The 2--8 keV to 1--2 keV hardness ratio map of G330.2+1.0 as obtained by
the {\it XMM-Newton} EPIC MOS1+MOS2. In the soft band (1--2 keV) map, the 1.4--1.6 and
1.7--1.8 keV bands are excluded to remove the bright instrumental lines at $E$ $\sim$ 1.5 
and 1.74 keV. For the purposes of display, each image has been binned into 15$^{\prime\prime}$ 
$\times$ 15$^{\prime\prime}$ pixels, and is adaptively smoothed to achieve a minimum S/N = 4. 
Green image contours are the 1--8 keV image of the SNR as shown in Fig.~\ref{fig:fig2}. 
Red contours are the 843 MHz radio image taken from the MOST Supernova Remnant Catalog 
\citep{wg96}. The angular resolution of the radio image is 43$^{\prime\prime}$. 
The position of the CCO J1601 is marked with a cross. The hard feature in the SW of
the CCO is marked with a white arrow. 
\label{fig:fig3}}
\end{figure}

\begin{figure}[]
\figurenum{4}
\centerline{\includegraphics[angle=0,width=0.70\textwidth]{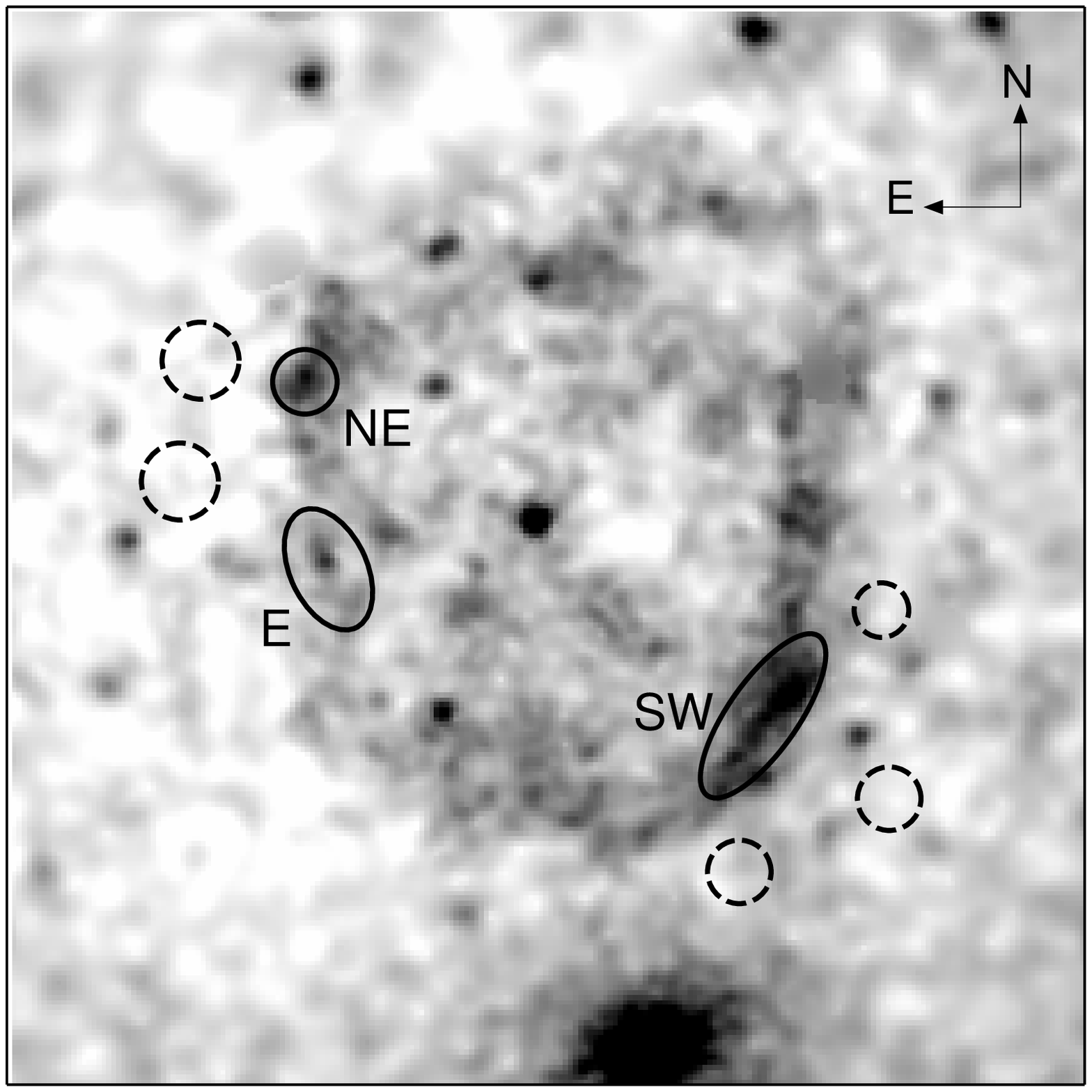}}
\figcaption[]{The broadband (1--8 keV) grey-scale image of G330.2+1.0 obtained by
{\it XMM-Newton} EPIC MOS1+MOS2. The image has been processed in the same way as
those in Fig.~\ref{fig:fig2}. SW, NE, and E regions of the SNR shell
are marked with solid lines. Background regions are marked with dashed circles. 
\label{fig:fig4}}
\end{figure}

\begin{figure}[]
\figurenum{5}
\centerline{\includegraphics[angle=0,width=0.9\textwidth]{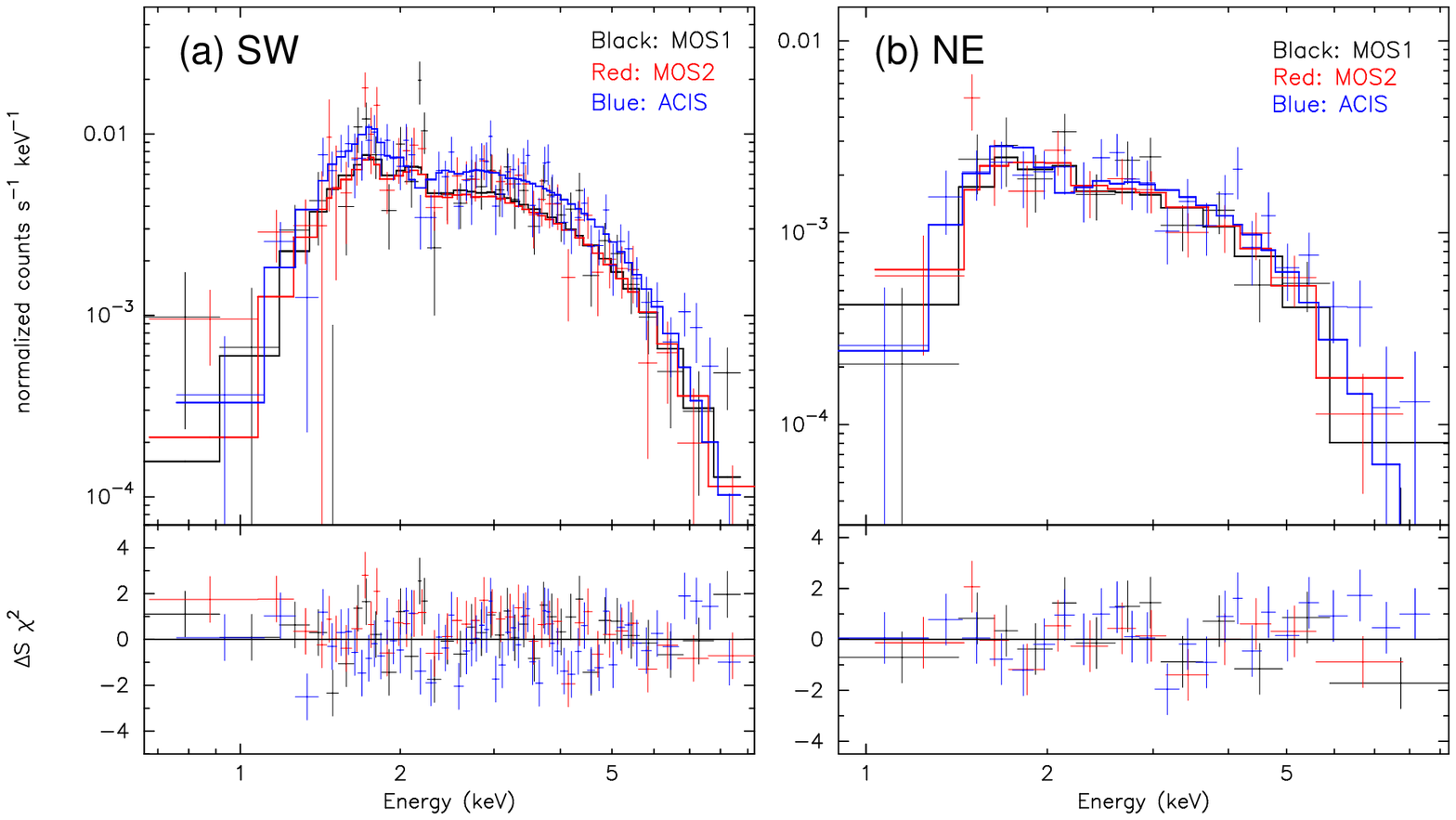}}
\figcaption[]{The X-ray spectrum from the shell of G330.2+1.0. (a) The SW, and
(b) the NE shell. The best-fit power law model for each regional spectrum is
overlaid. The lower panels are the residuals from the best-fit model.
\label{fig:fig5}}
\end{figure}

\begin{figure}[]
\figurenum{6}
\centerline{\includegraphics[angle=-90,width=0.7\textwidth]{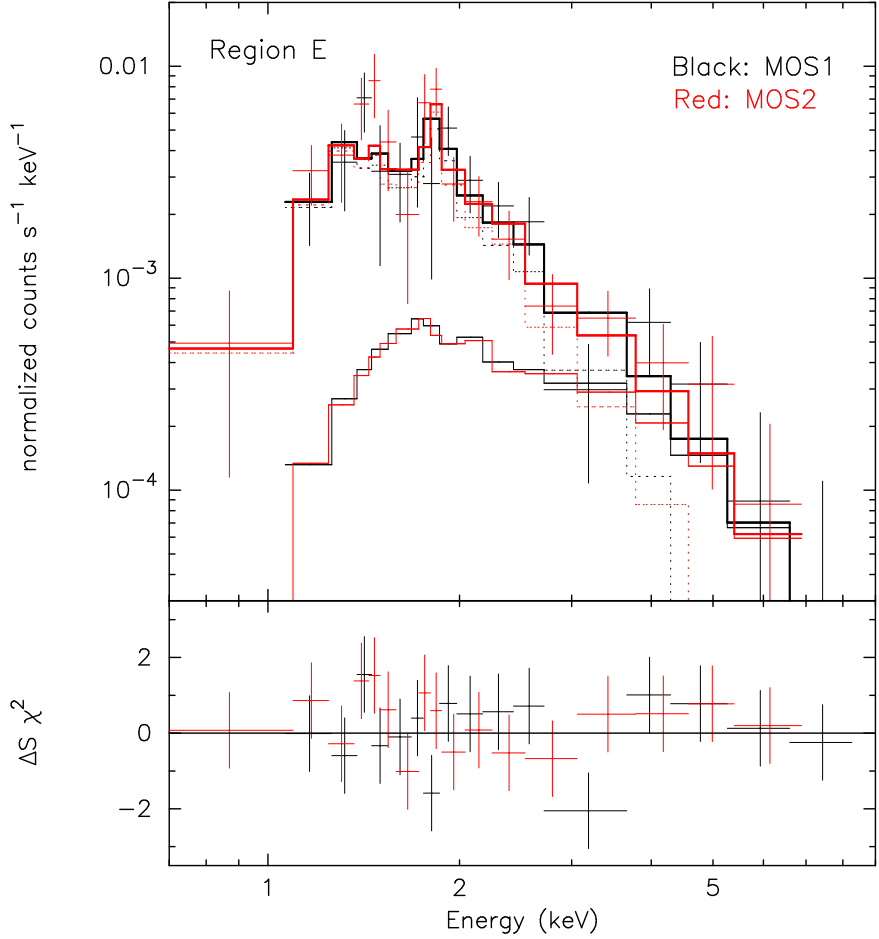}}
\figcaption[]{The X-ray spectrum from region E of G330.2+1.0. 
The best-fit two component model (thick solid lines) is overlaid. The dotted 
and the thin solid lines are the plane-shock ($kT$ = 0.7 keV) and the power law 
($\Gamma$ = 2.3) components, respectively. The lower panel is the residuals 
from the best-fit model.
\label{fig:fig6}}
\end{figure}

\begin{figure}[]
\figurenum{7}
\centerline{\includegraphics[angle=0,width=0.9\textwidth]{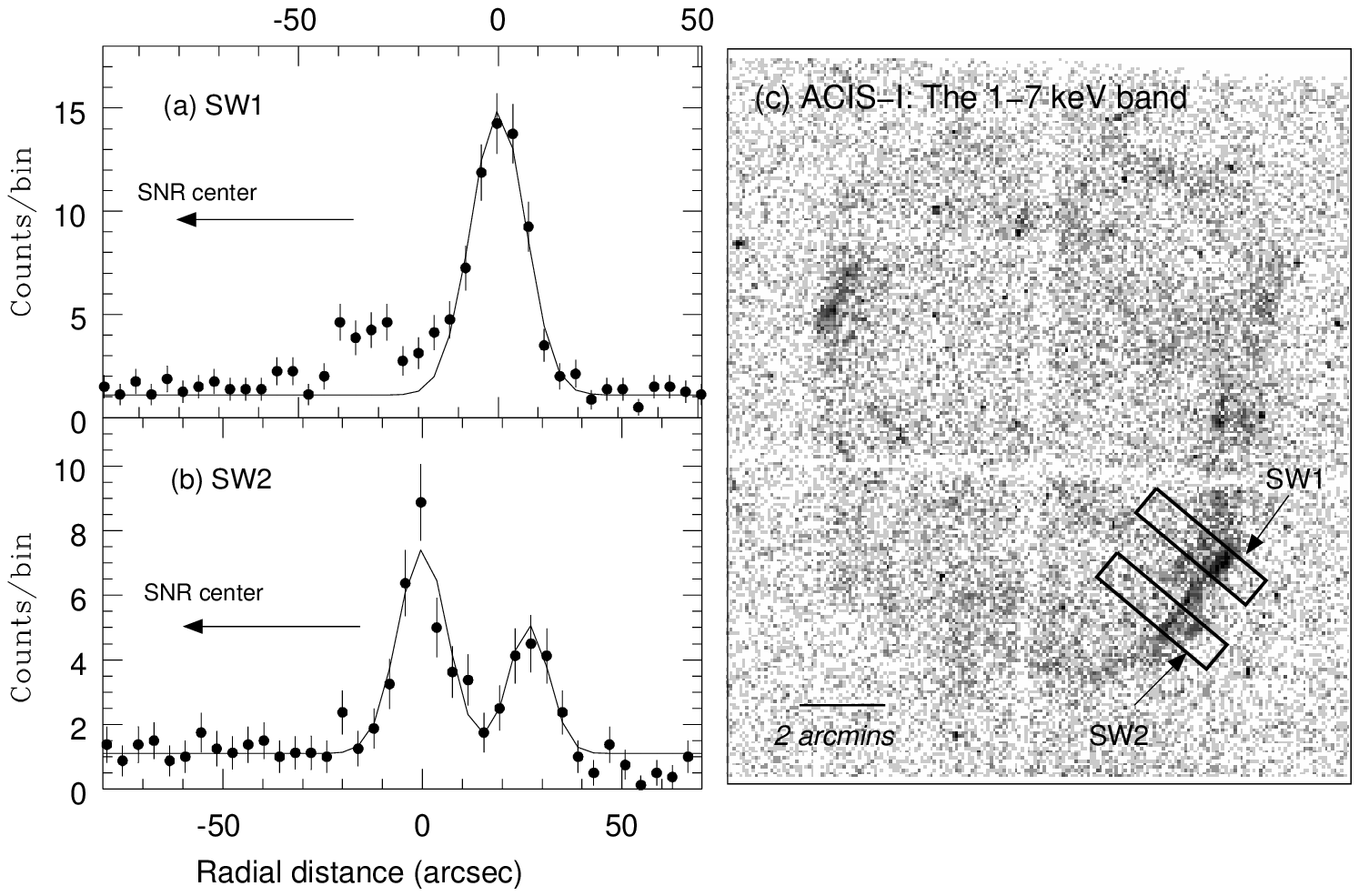}}
\figcaption[]{Radial intensity profiles of the SW shell of G330.2+1.0 obtained by the {\it Chandra}
data. (a) The bright northern parts (SW1) and (b) the faint southern parts (SW2) of the 
SW shell. (c) The 1--7 keV band {\it Chandra} ACIS-I image of G330.2+1.0. SW1 and SW2 
regions (180$^{\prime\prime}$ $\times$ 40$^{\prime\prime}$ for each region) are marked. 
The image has been binned into 4$^{\prime\prime}$ size pixels for the purposes of display.
In (a) and (b), each regional image has been binned into 8 $\times$ 8 pixels 
($\sim$4$^{\prime\prime}$ size), and then is averaged over 40$^{\prime\prime}$ column along 
the shell to produce a projected 1-D radial intensity profile. The best-fit Gaussian model 
(with a constant underlying background) is overlaid. In (a), the small intensity
bump just inside of the shell ($\sim$25$^{\prime\prime}$--45$^{\prime\prime}$ toward the 
SNR center) is excluded in the fit. 
\label{fig:fig7}}
\end{figure}

\end{document}